# Magnetic Magnetoelectric and Magnetoelastic Properties of new multiferroic material $NdFe_3(BO_3)_4$

A.M. Kadomtseva[1], A. V. Kuvardin[1], A. P. Pyatakov[1,2], A. K. Zvezdin[2], G. P. Vorob'ev[1],

Yu. F. Popov[1], L. N. Bezmaternykh[3]

[1]Moscow State University, Moscow, 119992 Russia

[2]General Physics Institute, Russian Academy of Sciences, Moscow, 119991 Russia

[3]Kirenskii Institute of Physics, Siberian Division, Russian Academy of Sciences, Akademgorodok, Krasnoyarsk,

660036 Russia

*Complex experimental and theoretical study of the magnetic, magnetoelectric, and magnetoelastic properties of neodymium iron borate $NdFe_3(BO_3)_4$ along various crystallographic directions have been carried out in strong pulsed magnetic fields up to 230 kOe in a temperature range of 4.2–50 K. It has been found that neodymium iron borate, as well as gadolinium iron borate, is a multiferroic. It has much larger (above 300 $\mu C/m^2$) electric polarization controlled by the magnetic field and giant quadratic magnetoelectric effect. The exchange field between the rare-earth and iron subsystems (~50 kOe) has been determined for the first time from experimental data. The theoretical analysis based on the magnetic symmetry and quantum properties of the Nd ion in the crystal provides an explanation of an unusual behavior of the magnetoelectric and magnetoelastic properties of neodymium iron borate in strong magnetic fields and correlation observed between them.*

## Introduction

In the family of rare-earth iron borates $RFe_3(BO_3)_4$, gadolinium iron borate $GdFe_3(BO_3)_4$ is the most thoroughly studied (the ground state of the $Gd^{3+}$ ion is $^8S_{7/2}$) [1–4]. It exhibits the structural phase transition at 156 K, antiferromagnetic ordering of the spins of $Fe^{3+}$ at $T_N$ = 38 K, and spin-reorientation transition at $T_R$ ~ 10 K. Magnetic field-induced phase transitions recently found in it and the effects of the magnetic field control over the electric polarization that accompany these phase transitions indicate multiferroic properties of this compound [5]. It was of interest to investigate the effect of the ground state and the ionic radius of the rare-earth ion in iron borates on their properties and phase transitions. In this end, complex investigation of various properties of a single crystal of neodymium iron borate $NdFe_3(BO_3)_4$ ( $T_N \sim 32$ K) were carried out. The $Nd^{3+}$ ion has the ion radius larger than the $Gd^{3+}$ ion and is a Kramers ion with the $^4I_{9/2}$ ground state. Neodymium iron borate $NdFe_3(BO_3)_4$, as well as gadolinium iron borate $GdFe_3(BO_3)_4$ has the rhombohedral structure with the space group R32 and the effect of the magnetic-field control over the electric polarization is possible in this compound as well as in $GdFe_3(BO_3)_4$ [5]. In order to compare the properties of $GdFe_3(BO_3)_4$ and $NdFe_3(BO_3)_4$, the magnetic, magnetoelectric, and magnetoelastic properties of single crystals of neodymium iron borate $NdFe_3(BO_3)_4$ in strong pulsed magnetic fields were investigated.

## Experimental results

The magnetization, electric polarization, and magnetostriction of the $NdFe_3(BO_3)_4$ single crystal were measured in strong magnetic fields up to 230 kOe for H||c and H⊥c for temperatures 4.2–50 K. Figure 1 shows the magnetization curves along the a- and c-axes of the $NdFe_3(BO_3)_4$ crystal for temperatures of 4.5, 15, 20, and 30 K (the a- axis is one of the three twofold axes perpendicular to the threefold c-axis). According to the measurements, when the field was applied along the c-axis, the magnetization was almost independent of temperature T. The magnetization curves along the a- and c-axes of the crystal coincided with each other and exhibited temperature dependence (see Fig. 1). The total magnetization of $NdFe_3(BO_3)_4$ consists of the slightly T-dependent contribution from the antiferromagnetic iron sublattice and the strongly T-dependent contribution from the rare-earth sublattice of the $Nd^{3+}$ ions:

$$\mathbf{M}(H,T) = \mathbf{M}_{Fe}(H) + \mathbf{M}_{Nd}(H,T) \qquad (1)$$

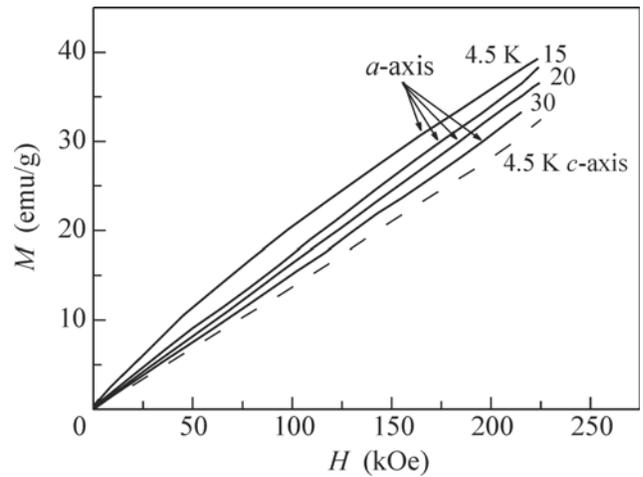

**Fig. 1.** Magnetization vs. the magnetic field directed along the (solid lines) *a*-axis and (dashed line) *c*-axis for the NdFe3(BO3)4 single crystal.

When the magnetic field was oriented along the a-axis of the crystal, longitudinal magnetostriction strains appeared at low temperatures 4.2–25 K and these deformations exhibited dome-like magnetic field dependence (see Fig. 2). In particular, at a temperature of 4.5K in a magnetic field of $H_{crit}\sim$10kOe, a positive jump was observed in the longitudinal magnetostriction ($\sim 10^{-5}$). As temperature increased, this jump decreased and vanished at 25K. When the magnetic field increased, longitudinal magnetostriction changed its sign and negative magnetostriction observed above 28 K was a quadratic function of the magnetic field. As shown below, a field of 50 kOe for which magnetostriction changes sign corresponds in the theoretical model to the exchange field generated by the antiferromagnetic iron subsystem on rare-earth ions. The stepwise appearance of large positive longitudinal electric polarization ($\sim$400 $\mu$C/m$^2$ at 4.5 K) was also found in the same temperature range 4.2–25 K for fields $H_{crit}\sim$10kOe parallel to the a-axis. When the field increased, this polarization decreased and changed its sign for the same field 50kOe as magnetostriction. However, in contrast to magnetostriction, electric polarization was saturated at a value of about − 320 $\mu$C/m$^2$ for fields above ~ 80 kOe (see Fig. 3). It is worth noting that the saturation electric polarization $P_a(H_a)$ in NdFe$_3$(BO$_3$)$_4$ is 30 times larger than the value observed for GdFe$_3$(BO$_3$)$_4$. The longitudinal magnetostriction and electric polarization for **H** || *c* for all temperatures 4.2–50 K are an order of magnitude lower than the respective values along the *a* axis and are quadratic functions of the magnetic field, which likely indicates to the orientation of the iron-ion spins in the *ab* plane. In the absence of the field, three equivalent directions corresponding to three twofold axes exist in the *ab* plane. The field applied in the plane reorients the magnetic moments of the iron sublattices in a field of $H \sim 10$ kOe in the direction perpendicular to the field. The fact that the longitudinal magnetostriction jumps for the magnetic field orientation along the *b* axis of the crystal (i.e., perpendicularly to one of the twofold axes) are an order of magnitude smaller [$\lambda_b(H_b) \sim 10^{-6}$] than the polarization jumps for **H** ∥ *a* possibly indicates that spins are predominantly oriented along the twofold axes.

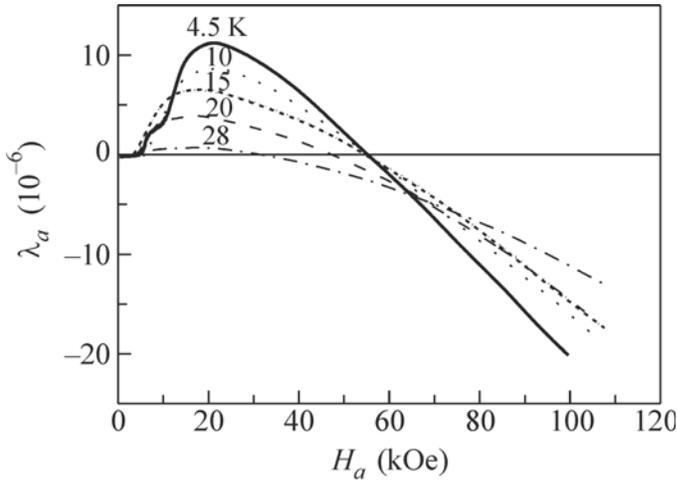

**Fig. 2.** Isotherms of the longitudinal magnetostriction vs. the magnetic field directed along the *a*-axis for the NdFe3(BO3)4 single crystal.

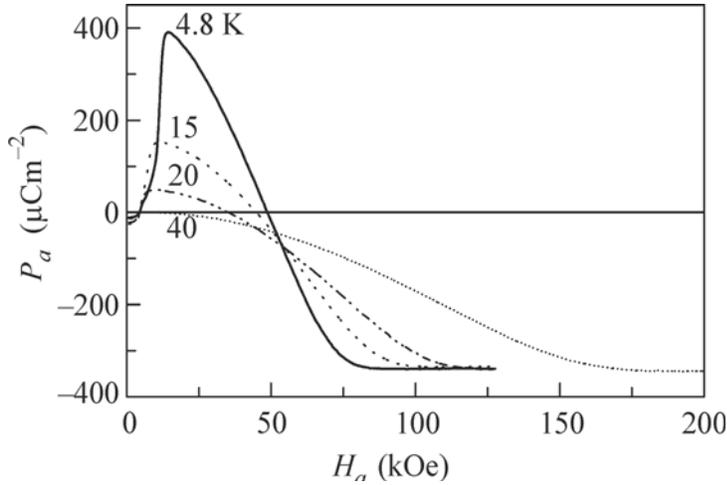

**Fig. 3.** Longitudinal electric polarization vs. the magnetic field directed along the - axis for the NdFe3(BO3)4 single crystal.

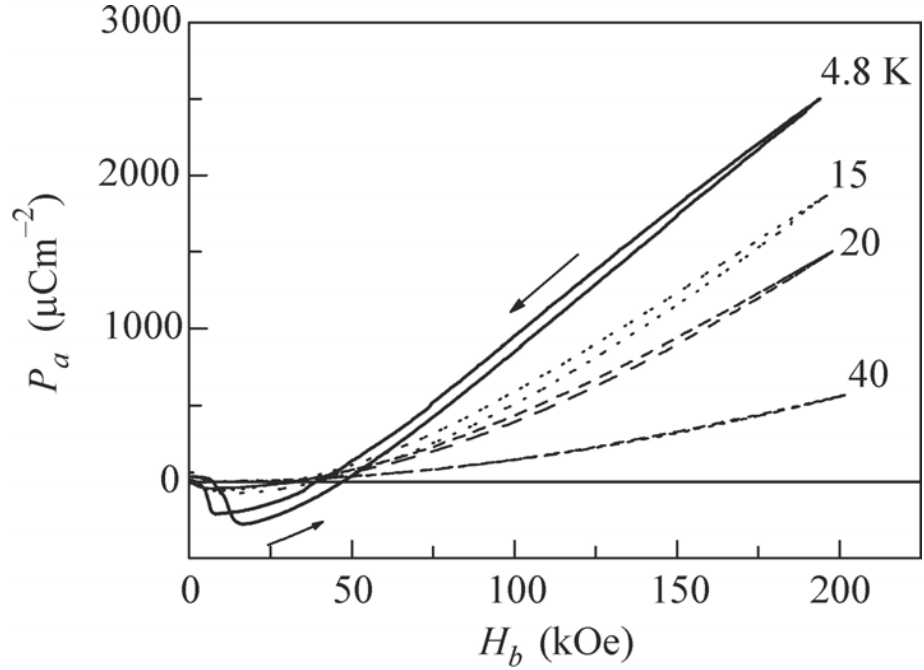

**Fig. 4.** Electric polarization along the *a* axis vs. the magnetic field directed along the *b* axis for the NdFe3(BO3)4 single crystal.

In contrast to the longitudinal electric polarization $P_a(H_b)$, the transverse electric polarization $P_a(H_a)$ in a strong magnetic field was not saturated and reached a large value of 2mC/m$^2$ in a field of 200 kOe at 4.5K (see Fig. 4).

**Magnetic symmetry, theory, and discussion**

To discuss the experimental results, we use the model of the magnetic structure of NdFe$_3$(BO$_3$)$_4$ that was previously considered in [2, 5] for gadolinium iron borate but supplement it with additional concepts concerning the subsystem of rare-earth ions.

The essence of this model is as follows. The subsystem of $Fe^{3+}$ ions is described by means of two magnetic sublattices antiferromagnetically coupled through the exchange field. Let $\mathbf{M}_1, \mathbf{M}_2$ be the magnetic moments of the iron sublattices; $\mathbf{M}_{Fe} = \mathbf{M}_1 + \mathbf{M}_2$, $\mathbf{L} = \mathbf{M}_1 - \mathbf{M}_2$ be the magnetization and antiferromagnetism vector of the iron ion subsystem, respectively; and $|\mathbf{M}_1| = |\mathbf{M}_2| \approx \frac{15}{2} \mu_B / f.u.$. The magnetic anisotropy of neodymium iron borate is such that the vector **L** lies in the basal plane (*ab* plane) of the crystal (for $H = 0$) in the entire temperature range from 0 to $T_N$ in contrast to $GdFe_3(BO_3)_4$, where the spins of irons are reoriented from the basal plane to the c axis at $T = 10$ K. According to the accepted concepts, this reorientation is caused by the interaction of Gd ions with the iron ion subsystem. For $NdFe_3(BO_3)_4$, the situation is opposite: Nd ions increase the stability of the basal plane for the spins of iron, i.e., increase the corresponding anisotropy constant confining the iron-ion spins in the *ab* plane. This is a characteristic property of $Nd^{3+}$ ions.

According to the symmetry of the crystalline structure of iron borates, the basal plane contains three easy magnetization axes coinciding with the twofold axes. Therefore, the crystal under equilibrium conditions is divided into three types of domains, the vector **L** oriented along the corresponding twofold axis in each of them. This circumstance should be taken into account when considering the magnetization processes in relatively weak fields (up to 10 kOe, as follows from experimental magnetoelectric and magnetostriction dependences).

For further analysis, it is natural to separately consider two magnetic field ranges: (i) relatively weak fields up to 10 kOe and (ii) strong fields above 10 kOe. In the former range, the process of magnetic and domain structure rearrangement in the sample takes place so that the vector *L* becomes uniform over the sample and directed perpendicularly to the external magnetic field. The latter remagnetization process is determined by both the rare-earth subsystem and the spin flip of the iron sublattice and it extends to the field range of about $10^6$ Oe; for this reason, for fields up to $(1-2) \times 10^5$ Oe, one may take

$$\mathbf{M}_{Fe} = \chi_\perp \mathbf{H}, \qquad (2)$$

here, $\chi_\perp$ is the transverse susceptibility of the iron sub-lattice (interaction with the rare-earth subsystem almost does not affect the spin-flip of the iron sublattices in neodymium iron borate).

Let us consider the system of $Nd^{3+}$ ions. As usual, we disregard the Nd–Nd exchange interaction (its energy 1 K); i.e., the Nd ion subsystem is treated as paramagnetic and is subjected to the exchange field created by iron ions. Another important action on Nd ions is provided by the crystal field that splits the main multiplet of these ions into five Kramers doublets.

The main multiplet of $Nd^{3+}$ is $^4I_{9/2}$ with S = 3/2, L = 6, and J= 9/2. According to the spectroscopic experiment reported in [6], the main doublet of the $^4I_{9/2}$ split multiplet is fairly well separated from excited doublets. For this reason, $Nd^{3+}$ ions can be treated in the one-doublet approximation over the entire temperature range (4.2–30 K) under investigation. The second important consequence of the optical experiment [6] is that the main doublet of $Nd^{3+}$ ions is split (splitting energy ~8.8 cm$^{-1}$ at 4.2 K and decreases continuously to zero at T = $T_N$). This means that $Nd^{3+}$ ions are magnetized, i.e., have magnetic moments for T < $T_N$. At the same time, the magnetization curves of $NdFe_3(BO_3)_4$ for H $\rightarrow 0$ approach zero for all orientations of the magnetic field. Taking into account these experimental data, we think that the $Nd^{3+}$ ion system should be divided into two subsystems each of which is characterized by the magnetic moment (per ion)

$$\mathbf{m}_{i,\alpha}(H,T) = \frac{g_\alpha \mu_B}{2} \frac{\mathbf{H}^i_{eff}}{H_{eff}} th\left(\frac{g_\alpha \mu_B H_{eff}}{2kT}\right) + \chi^{VV}_\alpha \mathbf{H}_{eff} \qquad (3)$$

Here, $g_\alpha$ is the effective g factor of the main doublet of the $Nd^{3+}$ ion; the subscript $\alpha = \perp$ and c specifies the orientation of the magnetic moment, $\perp$ (small effect of magnetic anisotropy in the plane on g is disregarded here) and c refer to the basal plane and c axis, respectively; $\mathbf{H}^i_{eff} = \pm\mathbf{H}_{exch} + \mathbf{H}$, where $\mathbf{H}_{exch}$ is the exchange field that is created by the iron sublattices and acts on rare-earth ions, the signs ± correspond to two antiferromagnetic sublattices, and the superscript i = 1, 2 specifies the rare-earth subsystems; and $\chi^{VV}_\alpha$ is the van Vleck susceptibility.

Formula (3) satisfies both the above optical experiment (single-doublet splitting in zero field) [6] and magnetic measurement data (vanishing of the total magnetization $\mathbf{M}_{Nd} = \mathbf{m}_1 + \mathbf{m}_2$ of the rare-earth sub-system in the $H \rightarrow 0$ limit). The value of the effective g-factor has not yet been determined with certainty. The magnetic moment of a free ion is determined as $\mu_0 = gJ\mu_B = \frac{8}{11} \times \frac{9}{2}\mu_B \approx 3.2\mu_B$. However, formula (3) corresponds to the single-doublet approximation. In this case, $g_\alpha$ is determined by the matrix elements of the magnetic moment operator $\langle 1|g\mu_B \mathbf{J}|2\rangle$ between the wave functions of the ground Kramers doublets $|1\rangle$ and $|2\rangle$. These wave functions are unknown, but according to the analysis of experimental data, the magnetic moment of the rare-earth ion for magnetization in the basal plane is equal to $g_\perp \mu_B \sim 1\mu_B$ as shown below, whereas the contribution from the rare-earth subsystem for magnetization along the c axis is equal to zero (i.e., $g_c \sim 0$), as indicated by the absence of the temperature dependence of $M(H \| c)$ (see Fig. 1).

In view of the above remarks, the total magnetization (per molecule) of the rare-earth ion system that is obtained by summing Eq. (3) with respect to $i$ is equal to

$$M_\perp^{ND} = \frac{g_\perp \mu_B}{2} \frac{H_\perp}{H_{eff}} th\left(\frac{g_\alpha \mu_B H_{eff}}{2kT}\right) + \chi_\perp^{VV} H_\perp, \quad (4\,a)$$

$$M_c^{ND} = \chi_c^{VV} H_c, \quad (4\,б)$$

where $H_\perp$ and $H_c$ are the in-plane and $\|$ c projection of the external field H, respectively.

It is assumed that the external field is large enough (>10 kOe) so that the antiferromagnetic structure is uniform and the spins in antiferromagnetic iron sublattices directed perpendicular to the external field H so that $H_{eff} = \sqrt{H_{exch}^2 + H_\perp^2}$ (see fig.5).

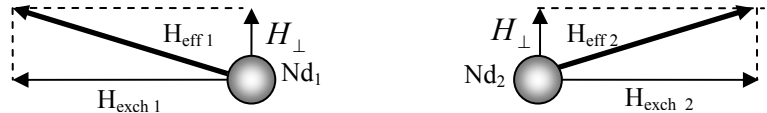

Fig. 5 Effective field acing on rare earth ions of two paramagnetic sublaticces $|H_{eff1}| = |H_{eff2}| = \sqrt{H_{exch}^2 + H_\perp^2}$

Taking the value Hexch ≈ 50 kOe for the exchange field between the rare-earth and iron sublattices (field in which the rare-earth contribution to electric polarization and magnetostriction changes sign, see Figs. 2 and 3) and using the splitting value $\Delta = g_{eff}\mu_B H = 8.8 см^{-1}$ from the optical experiment reported in [6], we estimate the effective g factor as $g_\perp \approx 2.7$. Disregarding van Vleck susceptibility $\chi_\alpha^{VV} \approx 0$ formulas (1), (2), and (4) with these data well reproduce the experimental curves plotted in Fig. 1. The dependence slope for $H \parallel c$ enables one to determine antiferromagnetic susceptibility $\chi_\perp \approx 1.4 \cdot 10^{-4} \frac{emu}{g \cdot Oe}$, which corresponds to the exchange field $H_E \sim 700$ kOe of the antiferromagnetic iron subsystem.

Let us analyze the magnetoelectric and magnetoelastic measurements. In order to mathematically describe the relation between the electric polarization $\mathbf{P}$, antiferromagnetism vector $\mathbf{L}$, iron sublattice magnetization $\mathbf{M}_{Fe}$, and the rare-earth subsystem magnetizations $\mathbf{m}_i$ we consider the transformation properties of the vectors $\mathbf{P}, \mathbf{L}, \mathbf{M}$ as well as of tensors, with respect to the space group as R32. It is worth noting that the magnetic unit cell of neodymium iron borate is twice as large as (in the c-axis direction) the crystal chemical unit cell; this circumstance presents the direct application of the traditional approach to the theory of antiferromagnetism that is based on magnetic and spatial symmetry [7–9] and implies that the crystal chemical unit cell coincides with the magnetic one. An obvious way to overcome this difficulty is to double the crystal chemical unit cell of neodymium iron borate and thereby to return to the traditional approach. Following this way, we reduce the space group R32, suggesting that the translations by the unit-cell constants in the basal plane and doubled translation along the c axis are equal to identity transformation E. The reduced group $\widetilde{G}_{32}$ is characterized by the following generators: ($E$) identity transformation, $C_3^+$ threefold axis parallel to the $c$-axis, $2_x^+$ one of the three twofold axes that is perpendicular to the $c$-axis and directed along the $a$ axis, and $T_1^-$ single-cell translation along the $c$ axis.

The signs ± denote exchange symmetry. In particular, the even symmetry operations $C_3^+$ and $2_x^+$ transform the antiferromagnetic sublattice to itself, whereas the odd operation $T_1^-$ transforms one antiferromagnetic sublattice to the other sublattice with the opposite magnetization direction.
Using these generators, one may construct the table of the transformation properties of necessary tensors for the space group $\widetilde{G}_{32}$. According to the table, the formulas for the components of the vector $\mathbf{P}$ have the form:

$$P_x = c_1 L_y L_z + c_2 \left(L_x^2 - L_y^2\right) + \frac{1}{2} \sum_{i=1}^{2} \left\{ c_3 \left(m_{ix}^2 - m_{iy}^2\right) + c_4 m_{iz} H_y + c_5 m_{iz} m_{iy} \right\} \quad (5a)$$

$$P_y = -c_1 L_x L_z - 2c_2 L_x L_y - \frac{1}{2}\sum_{i=1}^{2}\{2c_3 m_{ix} m_{iy} + c_4 m_{iz} H_x + c_5 m_{iz} m_{ix}\} \quad (5b)$$

$$P_z = c_6 L_x L_z (L_x^2 - 3L_y^2) + \frac{1}{2} c_7 \sum_{i=1}^{2} m_{ix} m_{iz}\{m_{ix}^2 - 3m_{iy}^2\}, \quad (5c)$$

where summation is performed over two paramagnetic subsystems of the rare-earth sublattice. Since the electric polarization in neodymium iron borate is larger than that in $GdFe_3(BO_3)_4$, the basic contribution to Eqs. (5) comes from the rare-earth subsystem, i.e., from terms determined by the constants $c_3$, $c_4$, $c_5$, and $c_7$. For this reason, terms with the coefficients $c_1$, $c_2$, and $c_6$ can be omitted in the first approximation in order to avoid lengthy formulas. The coefficients $c_3$, $c_4$, $c_5$, and $c_7$ in Eq. (5) are identical for both rare-earth sublattices, because all rare-earth ions occupy the identical crystallographic position and differ due only to the exchange field acting on them [see Eq. (3)].

|  | $E$ | $T_1$ | $C_3$ | $2_x$ | $L_i$ | $M_i$ | $L_i L_j$ | $M_i M_j$ | $P_i$ | $u_{ij}$ |
|---|---|---|---|---|---|---|---|---|---|---|
| $\Gamma_1$ | 1 | 1 | 1 | 1 |  |  | $L_x^2 + L_y^2$; $L_z^2$ | $M_x^2 + M_y^2$; $M_z^2$ |  | $u_{xx} + u_{yy}$; $u_{zz}$ |
| $\Gamma'_1$ | 1 | -1 | 1 | 1 |  |  |  |  |  |  |
| $\Gamma_2$ | 1 | 1 | 1 | -1 |  | $M_z$ |  |  | $P_z$ |  |
| $\Gamma'_2$ | 1 | -1 | 1 | -1 | $L_z$ |  |  |  |  |  |
| $\Gamma_3$ | 1 | 1 | $R$ | $\begin{pmatrix}1 & 0 \\ 0 & -1\end{pmatrix}$ |  | $\begin{pmatrix}M_x \\ M_y\end{pmatrix}$ | $\begin{pmatrix}L_y L_z \\ -L_x L_z\end{pmatrix}$ $\begin{pmatrix}L_x^2 - L_y^2 \\ -2L_x L_y\end{pmatrix}$ | $\begin{pmatrix}M_y M_z \\ -M_x M_z\end{pmatrix}$ $\begin{pmatrix}M_x^2 - M_y^2 \\ -2M_x M_y\end{pmatrix}$ | $\begin{pmatrix}P_x \\ P_y\end{pmatrix}$ | $\begin{pmatrix}u_{xx} - u_{yy} \\ -2u_{xy}\end{pmatrix}$; $\begin{pmatrix}u_{yz} \\ -u_{xz}\end{pmatrix}$ |
| $\Gamma'_3$ | 1 | -1 | $R$ | $\begin{pmatrix}1 & 0 \\ 0 & -1\end{pmatrix}$ | $\begin{pmatrix}L_x \\ L_y\end{pmatrix}$ |  |  |  |  |  |

Table 1. Transformation properties of the basic physical tensors in the group $\tilde{G}_{32}$; $R$ is the 120°-rotation matrix; $P_i$, $L_i$, and $M_i$ are the components of the electric polarization, antiferromagnetism, and magnetization, respectively; and $u_{ij}$ are the magnetostriction tensor components

We first consider the case $\mathbf{H} = (H_x, 0, 0)$. Suggesting that domains with the vectors $\mathbf{L}$ oriented along the three easy axes exist in the sample for $H_x = 0$, it is easy to verify that the polarization components averaged over these domains (it is assumed that they have the same total volume) <**P**i> vanish, where < > is the averaging symbol. As the field increases, the domains are shifted and the iron spins are reoriented to the *y* axis (as it usually occurs in antiferromagnets due to the anisotropy of susceptibility) so that a uniform state with **L** = (0, $L_y$, 0) and with corresponding maximum of $P_x$ is realized in the sample for $H_x \geq 10$ kOe. A jump in the $P_x$ component at $H \approx 10$ kOe is remarkable in Fig. 3. It is natural to attribute it to the spin flop in a domain with the initial state **L** = ($L_x$, 0, 0).

For $H > 10$ kOe, we have **L** = (0, ±$L$, 0), $m_{1x} = m_{2x}$, $m_{1y} = -m_{2y}$, $m_{1z} = m_{2z} \approx 0$, Hence, according to Eq. (3)–(5), (5)

$$P_x(H \| x) \sim \left( \frac{H^2 - H_{exch}^2}{H_{exch}^2 + H^2} \right). \tag{6}$$

At $H = H$exch $= 50$ kOe, the magnetization components along the *a*- and *b*-axes are equal to each other $|m_{iy}| = |m_{ix}|$ and the longitudinal polarization given by Eq. (6) vanishes, which is corroborated by the experimental data (see Fig. 3). The nature of the saturation of $P_x(H)$ isotherms is not completely clear. This is due probably to the nonlinear properties of the electric subsystem of the crystal.

We now consider the case **H** = (0, $H$, 0), where the iron ion spins are reoriented to the *x* axis. In this case, spin flop does not occur and spins are reoriented continuously in contrast to the preceding case, where a jump is observed. This behavior is demonstrated by the dependences shown in Fig. 4. For $H > 10$ kOe, we have **L** = (±$L$, 0, 0), $m_{1z} = m_{2z} \sim 0$, $m_{1y} = m_{2y}$, and $m_{1x} = -m_{2x}$. Hence, according to Eq. (3)–(5),

$$P_x(H \| y) = -P_x(H \| x), \tag{7}$$

where $P_x(H \| x)$ is given by Eq. (6). Formula (6) is qualitatively correct for the field dependence of the electric polarization $P_x(H \| y)$ for fields from 10 kOe to 70–80 kOe with the same value $H_{exch} \approx 50$ kOe. However, we point to certain systematic overestimation of experimental curves by Eq. (7). This discrepancy is discussed below.

As seen in Fig. 4, the behavior of the curve $P_x(H \| y)$ for $H > 80$ kOe differs significantly from the behavior of the curve $P_x(H \| x)$: in contrast to the latter, the former exhibits a strong, almost quadratic increase up to $H > 200$ kOe. This difference should be specially analyzed. Formulas (5)–(7), which are used for above analysis of the field dependence $P_i(H)$, do not take into account that the crystal symmetry changes from class 32 to class 2 [at **P** = ($P_x$, 0, 0)] when electric polarization appears in the sample. In particular, the magnetic susceptibility of rare-earth ions in class 2 is not a diagonal tensor as in class 32. Additional components $\chi_{23}$ and $\chi_{32}$ appear in the tensor $\chi_{ik}$ (in the coordinate system, where the $x$ axis is parallel to the twofold axis). This means that the $z$ component $m_{iz} = \chi_{32}H_y$ of the magnetization of Nd ions appears for $\mathbf{H} = (0, H, 0)$, which leads to the appearance of an additional term in $P_x(H \| y)$:

$$\Delta P_x(H \| y) = c_4 m_z H_y + c_5 m_z m_y \sim H_y^2, \quad (8)$$

This circumstance explains the increase in polarization in strong fields, as well as the above systematic overestimation of the experimental curves by Eq. (7) in moderate fields (the comparatively small disregarded contribution from the iron sublattice, which is proportional to $L_x^2 - L_y^2$, affects in the same direction).

It is worth noting that $m_{iz} = 0$ in the magnetic field $\mathbf{H} = (H_x, 0, 0)$ in class 2 according to the symmetry of the tensor $\chi_{ik}$. Therefore, additional terms proportional to $m_z$ do not appear in electric polarization for such field orientation.

According to the table, the expression for the magnetoelastic energy contains the terms

$$f_{M-Elastic} = ... + c_1(u_{xx} - u_{yy})(L_x^2 - L_y^2) + c_2(u_{xx} - u_{yy})\sum_{i=1}^{2}\{m_{ix}^2 - m_{iy}^2\} + ... \quad (8)$$

According to this expression, terms corresponding to longitudinal magnetostriction $u_{xx} - u_{yy}$ are proportional to the difference $m_x^2 - m_y^2$ similar to the $x$ component of the polarization [see Eq. (5a)]. This property explains the correlation observed in the experimental dependences for the longitudinal polarization (see Fig. 3) and longitudinal magnetostriction (see Fig. 2): the effect changes sign for the same field ~ 50 kOe, which is equal to the $f$–$d$ exchange field.

The quadratic field dependence of the electric polarization and magnetostriction in the case $H \| c$ is qualitatively explained by the spin flip of the iron sublattices in the field [see Eq. (2)].

Thus, the investigations reported above indicate that neodymium iron borate, as well as gadolinium iron borate, is a multiferroic, but it has much larger electric polarization (above 300 $\mu C/m^2$) controlled by the magnetic field and giant quadratic magnetoelectric effect (on the order of $10^{-17}$ s/A). On the basis of the theoretical model describing the magnetic properties of the neodymium ion in the crystal, the experimental magnetization curves of neodymium iron borate are explained and the $f$–$d$ exchange field (50 kOe), as well as the $g$ factor for the neodymium ion, is estimated for the directions in the basal plane $g_\perp \sim 3$ and the principal axis $g_c \sim 0$. The extraordinary behavior of magnetoelectric and magnetoelastic properties and correlation observed between them are explained in the framework of the symmetry approach. The difference between the field dependences for the longitudinal $P_a(H_a)$ and transverse $P_a(H_b)$ electric polarizations are also explained.

We are grateful to M.N. Popova and A.N. Vasil'ev for their constant interest in this work. This work was supported by the Russian Foundation for Basic Research (project nos. 04-02-16592-a, 05-02-16997-a, and 04-02-81046-Bel2004-a) and in part by the "Dynasty" Foundation. A.P.P. acknowledges the support of the Council of the President of the Russian Federation for Support of Young Scientists and Leading Scientific Schools (project no. MK-3764.2005).